\documentclass[a4paper,fleqn,usenatbib]{mnras}
\usepackage[T1]{fontenc}
\usepackage{ae,aecompl}

\usepackage{amsmath}
\usepackage{graphicx}
\usepackage{lscape}
\usepackage{indentfirst}

\usepackage{amssymb}
\usepackage{color}
\usepackage[flushleft]{threeparttable}

\newcommand {\bc}{\begin {center}}
\newcommand {\ec}{\end {center}}
\newcommand {\be}{\begin {equation}}
\newcommand {\ee}{\end {equation}}
\newcommand {\beq}{\begin {eqnarray}}
\newcommand {\eeq}{\end {eqnarray}}

\newcommand {\ergs}{{\rm erg\ \rm s^{-1}}}
\newcommand {\comment}[1]{}

\def\M {{\rm M}}

\def\intl {\int\limits}
\def\lbar {\lambda\hskip-5pt\raise3pt\hbox {--}}
\def\lbr {\lambda\raise2pt\hbox {\hskip-4pt{\scriptsize --}}_\C}

\renewcommand{\d}{{\rm d}}

\def\cI {{\mathcal I}}

\def\ch {{\rm ch}}

\renewcommand{\d}{{\rm d}}

\title[Pairs in accretion columns of X-ray pulsars]
{Electron-positron pairs in hot plasma of accretion column in bright X-ray pulsars}
\author[A.~A.~Mushtukov et al.] 
{Alexander~A.~Mushtukov,$^{1,2,3}$\thanks{E-mail: al.mushtukov@gmail.com (AAM)}  
Igor~S.~Ognev,$^{3}$ and
Dmitrij~I.~Nagirner$^{4}$ \\ 
$^1$ Leiden Observatory, Leiden University, NL-2300RA Leiden, The Netherlands \\
$^2$ Space Research Institute of the Russian Academy of Sciences, Profsoyuznaya Str. 84/32, Moscow 117997, Russia \\
$^3$ P. G. Demidov Yaroslavl State University, Sovietskaya 14, 150003 Yaroslavl, Russia \\
$^4$ Sobolev Astronomical Institute, Saint Petersburg State University, Saint-Petersburg 198504, Russia \\  
} 

\pubyear{2018}

\begin{document}
\label{firstpage}
\pagerange{\pageref{firstpage}--\pageref{lastpage}}
\maketitle

\begin{abstract}
The luminosity of X-ray pulsars powered by accretion onto magnetized neutron stars covers a wide range over a few orders of magnitude.
The brightest X-ray pulsars recently discovered as pulsating ultraluminous X-ray sources reach accretion luminosity above $10^{40}\,\ergs$ which exceeds the Eddington value more than by a factor of ten.
Most of the energy is released within small regions in the vicinity of magnetic poles of accreting neutron star - in accretion columns. 
Because of the extreme energy release within a small volume accretion columns of bright X-ray pulsars are ones of the hottest places in the Universe, where the internal temperature can exceed 100 keV.
Under these conditions, the processes of creation and annihilation of electron-positron pairs can be influential but have been largely neglected in theoretical models of accretion columns. 
In this letter, we investigate properties of a gas of electron-positron pairs under physical conditions typical for accretion columns. We argue that the process of pairs creation can crucially influence both the dynamics of the accretion process and internal structure of accretion column limiting its internal temperature, dropping the local Eddington flux and increasing the gas pressure. 
\end{abstract}



\section{Introduction}

The majority of accreting X-ray binaries is represented by the systems hosting neutron stars (NSs).
Highly magnetized NSs form a special class of objects among X-ray binaries powered by accretion - X-ray pulsars (XRPs, see e.g. \citealt{2015A&ARv..23....2W}).
Typical magnetic field strength at the NS surface in XRPs is $\gtrsim 10^{12}\,{\rm G}$, which is confirmed independently by a number of different methods: 
detection of cyclotron lines (see \citealt{2019A&A...622A..61S} for review), 
transitions into the ``propeller" state \citep{2016A&A...593A..16T,2017ApJ...834..209L},
detection of spin-up and spin-down effects \citep{2017PASJ...69..100S}.
Detected luminosities of XRPs cover a few orders of magnitude from $10^{33}\,\ergs$ and up to $10^{41}\,\ergs$,
where the brightest pulsars belong to the recently discovered class of pulsating ultraluminous X-ray sources (ULXs,  \citealt{2014Natur.514..202B,2017Sci...355..817I}).
The theoretical explanation of ULX pulsars is still under debates, but the most of the theories agree that the extreme accretion onto magnetized NS results in the formation of accretion columns above the stellar surface, where the matter is confined by a strong magnetic field and produce luminosity well above the Eddington limit \citep{1976MNRAS.175..395B,1981A&A....93..255W,2015MNRAS.454.2539M}.

A strong magnetic field of a NS in XRP modifies both the geometry of accretion flow (see Chapter 6 in \citealt{2002apa..book.....F}) and basic properties of matter \citep{2006RPPh...69.2631H}.
Because of a strong magnetic field, the accreting material reaches NS surface in small regions (the typical area is $\sim 10^{10}\,{\rm cm^2}$) in the vicinity of NS magnetic poles.
At extremely high mass accretion rates the radiation pressure is high enough to stop accretion flow above NS surface in radiation dominated shock \citep{1982SvAL....8..330L,2015MNRAS.447.1847M}. 
Below the shock, the matter slowly settles to the stellar surface converting its gravitational and kinetic energy into emission in X-ray energy band. 
A strong energy release within a small region results in extreme temperatures, which are typically about a few keV \citep{1981A&A....93..255W,2015MNRAS.454.2539M} and can be as high as about $1\,{\rm MeV}$ in the case of ULX pulsars \citep{2018MNRAS.476.2867M}.

High temperatures in regions of energy release of XRPs result in specific processes which might shape properties of XRPs at extreme mass accretion rates and thus have to be taken into account in theoretical models. 
In particular, high temperatures result in the creation of electron-positron pairs and possibly strong emission of neutrino due to pair annihilation and/or cyclotron neutrino emission \citep{1992PhRvD..46.4133K,1994A&AT....4..283K,2018MNRAS.476.2867M}.

In this papers, we investigate properties of electron-positron gas under conditions of a strong magnetic field and high temperatures typical for internal regions of optically thick accretion columns.
Under the assumption of thermodynamic equilibrium, we calculate the number density of the pairs (Section \ref{sec:N}) and the gas pressure along magnetic field lines (Section \ref{sec:P}).
We discuss the influence of the pair creation process on internal temperature and dynamics of the accretion process.

\section{Electron-positron pairs in accretion column}

The accretion columns in bright XRPs are confined by a strong magnetic field and supported by the internal radiation and gas pressure \citep{1976MNRAS.175..395B}. 
The conditions of matter inside the columns are determined by the local mass density and temperature.
The main mechanism of opacity in accretion columns is Compton scattering of photons by electrons/positrons influenced by a strong magnetic field.
Because the accretion column is optically thick due to the scattering, the photons created in the accretion flow undergo a number of scatterings before they leave the column.
A fraction of radiation is truly absorbed while the photons diffuse towards the edges of the accretion channel.

Let consider a case of fully ionized hydrogen plasma.
A number density of protons $n_{\rm p}$ is determined by the mass accretion rate $\dot{M}$, cross section of the accretion channel $S$ and local velocity $v=\beta c$ of accretion flow 
\beq\label{eq:electron_ini}
n_{\rm p}\approx \frac{\dot{M}}{2 S v m_{\rm p}}\simeq 10^{23}\frac{\dot{M}_{20}}{S_{10}\beta}\quad {\rm cm^{-3}},
\eeq
where $\dot{M}_{20}=\dot{M}/10^{20}\,{\rm g\,s^{-1}}$ and $S_{10}=S/10^{10}\,{\rm cm^2}$.
Because the typical geometrical thickness of accretion column is of order of $d\sim10^4\,{\rm cm}$, the optical thickness of the column across magnetic field lines due to the scattering can be estimated as 
$
\tau_{\perp}\gtrsim  n_{\rm p} \sigma_\perp d 
\simeq 6.7\times 10^2\,{\dot{M}_{20}}{S_{20}^{-1}\beta^{-1}}\left({\sigma_{\perp}}/{\sigma_{\rm T}}\right),
$
where $\sigma_{\perp}$ is a scattering cross section across $B$-field lines and $\sigma_{\rm T}$ is the Thomson cross section.
The number of scattering experienced by photon in accretion column is
\beq
N_{\rm sc}\sim \tau_{\perp}^2\approx 4.4\times 10^5\,\frac{\dot{M}^2_{20}}{S^2_{20}\beta^2}
\left(\frac{\sigma_\perp}{\sigma_{\rm T}}\right)^2.
\eeq
The absorption cross section can be roughly estimated as 
$\sigma_{\rm a}\simeq 23.7\,\sigma_{\rm T}Z^4 (E/1\,{\rm keV})^{-3}$, where $Z$ is the atomic number, $E$ is a photon energy \citep{1957qmot.book.....B}.
The accreting material is slowing down from a free-fall velocity $v_{\rm ff}$ to about $v_{\rm ff}/10$ in the radiation dominated shock at the top of accretion column
Thus, the velocity below the shock region can be estimated from above as $v<0.05c$ and a typical number of absorption events corresponding to a given number of scatterings can be estimated as
\beq\label{eq:Nabs}
N_{\rm a}\sim N_{\rm sc}\frac{\sigma_{\rm a}}{\sigma_\perp}
\approx 30\,\frac{\dot{M}^2_{20}\,Z^4}{S^2_{20}(\beta/0.05)^2}\left(\frac{\sigma_\perp}{\sigma_{\rm T}}\right) 
\left(\frac{E}{511\,{\rm keV}}\right)^{-3}.
\eeq
The estimated number of absorption events will be even larger if one will account for the actual chemical composition of accreting material, which is affected by nuclear reactions in accretion column
In particular, for the case of carbon dominated material $Z\simeq 6$ and an estimated number of absorption events (\ref{eq:Nabs}) increases by a factor of $\sim 10^3$.
Thus, the photons emitted in the accretion column will be likely absorbed on their way to the edges of the accretion channel in the case of sufficiently large mass accretion rates.
Under these conditions, the gas of electron/positron pair can be considered to be in thermodynamic equilibrium \citep{1971SvA....15...17B}.
It worth to note that the outer layers of the accretion column might be far from thermodynamic equilibrium and more detailed analyses accounting for detailed balance between a number of processes is required (see e.g. \citealt{1985MNRAS.212..523G}).

Another essential condition for establishing of the equilibrium concentrations and distributions of electrons and positrons is sufficiently long dynamical time scale in accretion column: the dynamical time scale has to be longer than the time scale required for equilibrium establishing.
Because the annihilation cross section is of order of $\sigma_{\rm ann}\sim 10^{-24}\,{\rm cm^2}$ (see e.g. \citealt{1980ApJ...238..296D}) and typical velocity of electrons and positrons are of order of speed of light at temperatures $>50\,{\rm keV}$, the time scale of equilibrium establishing can be estimated as 
$$t_{\rm eq}\lesssim \frac{1}{\sigma_{\rm ann}n_{\rm p} c}\lesssim 3\times 10^{-10}\,\frac{S_{10}\beta}{\dot{M}_{20}}\,\,{\rm s},$$
while the dynamical time scale in the column can be estimated as
$$t_{\rm dyn}\sim \frac{R}{\beta c}\approx 3\times 10^{-5}\,\beta^{-1}\,\,{\rm s}.$$
Thus, the time scale of equilibrium establishing is much smaller than the dynamical time scale ($t_{\rm eq}\ll t_{\rm dyn}$) and pairs in the central parts of accretion column can be considered to be in the equilibrium.

\subsection{Number density of electrons and positrons}
\label{sec:N}

At low temperature regime, the number density of electrons is similar to the number density of protons $n_{-}\simeq n_{\rm p}$, while
at high temperature regime, the creation of electron-positron pairs becomes essential and the number density of leptons (both electrons and positrons) can significantly exceed the number density of protons.
The number density of electrons $n_{-}$ and positrons $n_{+}$ in the equilibrium are given by \citep{1968PhRv..173.1220C,1992PhRvD..46.4133K}
\beq \label{eq:EE_NumberDensity}
n_{\mp}&=&\left(\frac{mc}{\hbar}\right)^3 \frac{b}{4\pi^2}\sum\limits_{n=0}^{\infty}g_n 
\int\limits_{-\infty}^{\infty}\d p_{\rm z}\,f_{\mp}\\
&\approx & 4.415\times 10^{29}\,b \sum\limits_{n=0}^{\infty}g_n 
\int\limits_{-\infty}^{\infty}\d p_{\rm z}\,f_{\mp}\quad{\rm cm^{-3}} \nonumber,
\eeq
where $b=B/B_{\rm crit}$ is dimensionless magnetic field strength, $B_{\rm crit}\equiv m_{\rm e}^2 c^2/e\hslash \simeq 4.412\times 10^{13}\,{\rm G}$,
the distribution functions of particles
\beq \label{eq:ElectroNeutral}
f_{\mp}=\left( \exp\left[ \frac{E_n(p_{\rm z})\mp\mu}{t} \right]+1 \right)^{-1},
\eeq
dimensionless energy $E_n(p_{\rm z})=(1+p_{\rm z}^2+2bn)^{1/2}$ of electron/positron at $n$th Landau level,
dimensionless temperature $t\equiv T/( m_{\rm e}c^2)$, 
$g_n$ is the spin degeneracy of Landau level ($g_0=1$ and $g_n=2$ for $n\ge 1$),
and 
the dimensionless (in units of $m_{\rm e}c^2$) chemical potential $\mu$ is determined by the condition of electro-neutrality 
$ n_{-}-n_{+}=n_{\rm p}$.

\begin{figure}
\centering 
\includegraphics[width=8.5cm]{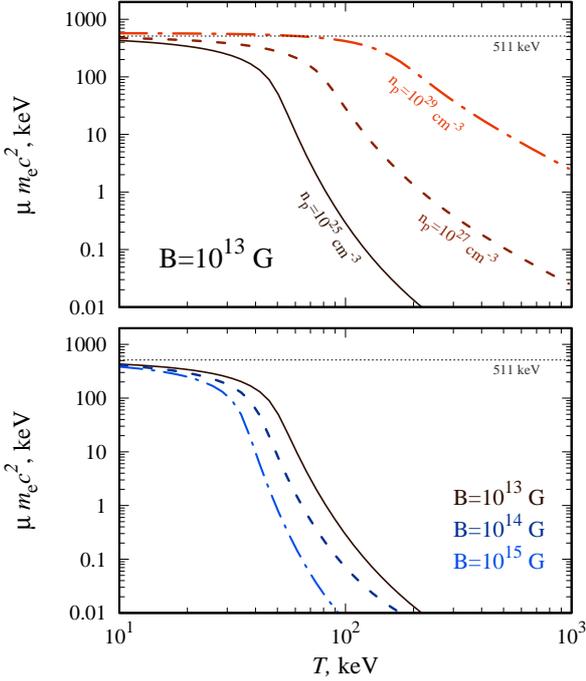} 	
\caption{The chemical potential of electrons calculated as a function of temperature for different number densities of protons at fixed magnetic field strength $10^{13}\,{\rm G}$ (top panel), and for different magnetic field strength at fixed number density of protons $n_{\rm p}=10^{25}\,{\rm cm^{-3}}$ (bottom panel).}
\label{pic:sc_mu01}
\end{figure}

\begin{figure}
\centering 
\includegraphics[width=8.5cm]{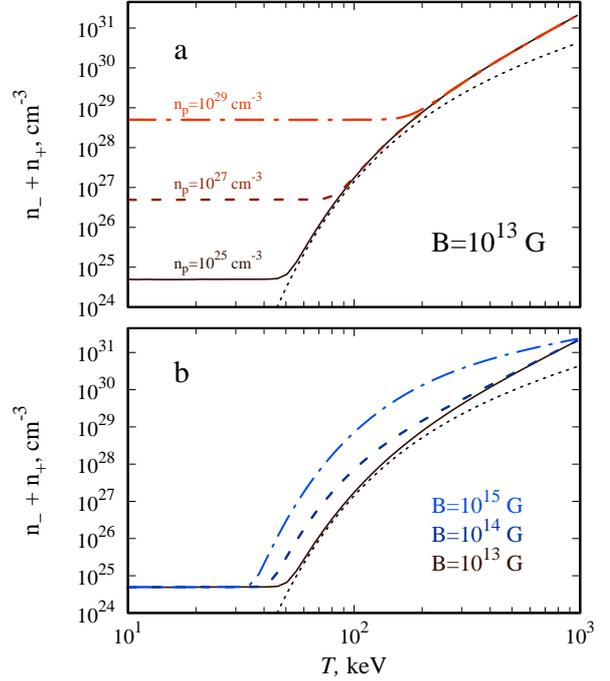} 	
\caption{Number density of electrons and positrons as a function of temperature calculated for 
(a) different number densities of protons: $10^{25}$ (black solid line), $10^{27}$ (brown dashed lile) and $10^{29}\,{\rm cm^{-3}}$ (red dashed-dotted line),
and (b) different magnetic field strength: $10^{13}$ (black solid line), $10^{14}$ (dark-blue dashed line) and $10^{15}\,{\rm G}$ (blue dashed-dotted line).
Black dotted line represents approximation given by (\ref{eq:NeApp}).}
\label{pic:sc_Ne01}
\end{figure}

In the limit of zero magnetic field and relatively low temperatures ($T\ll m_{\rm e}c^2$) the number densities of electrons and positrons can be estimated as \citep{1971reas.book.....Z}
\beq\label{eq:NeApp}
n_{+}\simeq n_{-}&\simeq & \frac{1}{\pi^2}\left( \frac{m_{\rm e}c}{\hbar} \right)^3 e^{-1/t}\,t^{3/2}\nonumber \\
&\simeq & 1.8\times 10^{30}\, e^{-1/t}\,t^{3/2} \quad {\rm cm^{-3}},
\eeq
while at high temperatures ($T\gg m_{\rm e}c^2$) the approximations is given by
\beq\label{eq:NeApp2}
n_{+}\simeq n_{-}\simeq \frac{1.82}{\pi^2}\left( \frac{m_{\rm e}c}{\hbar} \right)^3\,t^{3}
\simeq 3.2\times 10^{30}\,t^{3} \quad {\rm cm^{-3}}.
\eeq
The calculations with the accurate equations (\ref{eq:EE_NumberDensity}) and (\ref{eq:ElectroNeutral}) give a result which depends both on number density of protons and magnetic field strength (see Fig.\,\ref{pic:sc_mu01},\ref{pic:sc_Ne01}).
Influence of a strong magnetic field on the chemical potential and number density of electron-positron pairs becomes valuable at $B\gtrsim 10^{13}\,{\rm G}$. At lower $B$-field strength and temperature $T\gtrsim 10\,{\rm keV}$, the chemical potential and number density of the pairs can be calculated in approximation of zero field strength.
In the case of low number density of protons $n_{\rm p}$, the contribution of created electron-positron pairs to the total number density of leptons is dominant at temperatures of a few tens of keV already (see Fig.\,\ref{pic:sc_Ne01}a), which is typical temperature for accretion column interior \citep{2015MNRAS.454.2539M}. 
Note, that the stronger the magnetic field, the larger the number density of electron-positron pairs (see Fig.\,\ref{pic:sc_Ne01}b).

\begin{figure}
\centering 
\includegraphics[width=8.5cm]{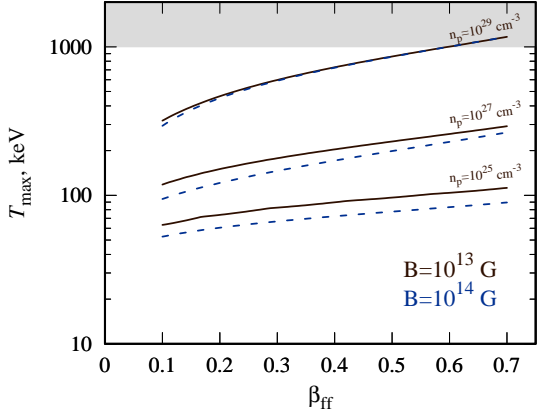} 	
\caption{The maximal temperature achievable in accretion column as a function of local dimensionless free fall velocity $\beta_{\rm ff}$.
Different curves are given for various number density of protons in the accretion flow: $n_{\rm p}=10^{25},\,10^{27}$ and $10^{29}\,{\rm cm^{-3}}$ (down up). Black solid and dark-blue dashed line correspond to local magnetic field strength of $10^{13}$ and $10^{14}\,{\rm G}$ respectively.
At temperatures above $\sim 1\,{\rm MeV}$ (grey area) the energy losses are strongly affected by neutrino emission.
}
\label{pic:sc_Tmax}
\end{figure}

The fast increase of a number density of electron-positron pairs with temperature requires that a fraction of energy of accreting material is going into a process of pair creation. 
Because the energy budget of the accretion process is limited by the mass accretion rate and compactness of a NS, the process of pairs creation limits the increase of internal temperature in the accretion column.
The upper limit of a number density of the pairs can be obtained from the assumption that the energy of accretion flow is going entirely in a pair creation. 
The total energy budget per one baryon due to the accretion process is determined by local free-fall velocity: $(\gamma_{\rm ff}-\gamma)m_{\rm p}c^2$ where $\gamma_{\rm ff}=(1-\beta_{\rm ff}^2)^{-1/2}$ is the Lorentz factor due to a local free-fall velocity and $\gamma$ is actual Lorentz factor of accreting material at a given height above stellar surface.
Then the number density of the pairs, which is similar to the number density of positrons $n_{+}$ in accretion flow, can be limited from above by 
\beq
(\gamma_{\rm ff}-\gamma)m_{\rm p}n_{\rm p}>2m_{\rm e}n_{+}. 
\eeq 
Thus, a conservative upper limit for the number density of pairs is 
\beq
n_{\rm +,max}\le \frac{n_{\rm p}}{2}\frac{m_{\rm p}}{m_{\rm e}}(\gamma_{\rm ff}-1). 
\eeq
The maximal number density of pairs determines the maximal temperature $T_{\rm max}$ achievable at a given local free-fall velocity $\beta_{\rm ff}$ and number density of protons (see Fig.\,\ref{pic:sc_Tmax}). 
One can see that the temperatures of a few hundred keV in accretion channel are achievable only in the case of sufficiently high densities (see Fig.\,\ref{pic:sc_Tmax}). 

\subsection{Gas and radiation pressure in accretion column}
\label{sec:P}

\begin{figure}
\centering 
\includegraphics[width=8.5cm]{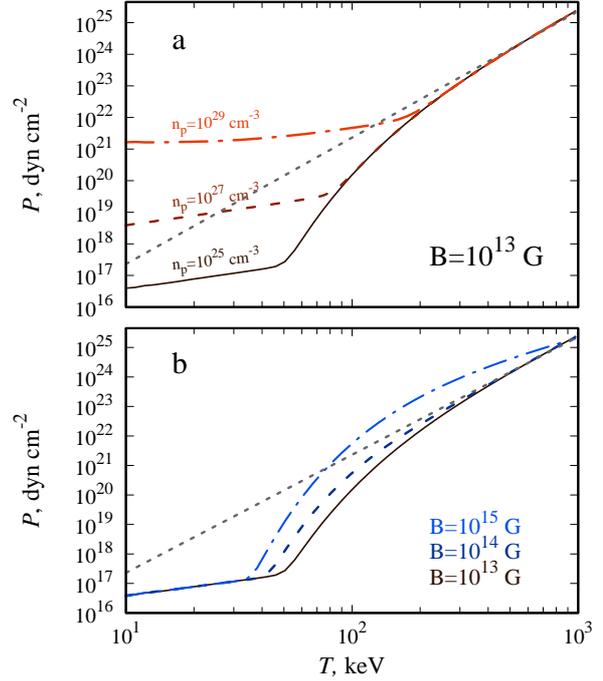} 	
\caption{The pressure of electron-positron gas along magnetic field lines calculated as a function of temperature for 
(a) different number densities of protons: $10^{25}$ (black solid line), $10^{27}$ (brown dashed lile) and $10^{29}\,{\rm cm^{-3}}$ (red dashed-dotted line),
and (b) different magnetic field strength: $10^{13}$ (black solid line), $10^{14}$ (dark-blue dashed line) and $10^{15}\,{\rm G}$ (blue dashed-dotted line).
Black dotted line represents radiation pressure calculated according to (\ref{eq:RadPressure}). }
\label{pic:sc_Pressure01}
\end{figure}

The gas pressure along $z$-direction due to the electron and positrons can be derived from the distribution function of the particles:
\beq\label{eq:eePressure1}
P_{\rm z, e e^+}=\frac{m_{\rm e}^4 c^5}{\hbar^3}\frac{b}{4\pi}
\sum\limits_{n=1}^{\infty}g_n \int\limits_{-\infty}^{\infty}\d p_{\rm z}\,\beta_{\rm z}p_{\rm z} [f_{-}+f_{+}]
\eeq
where the dimensionless velocity along $z$ axis  $\beta_{\rm z}=p_{\rm z}\gamma^{-1}=p_{\rm z}(1+p_{\rm z}^2+2bn)^{-1/2}$.
Expression (\ref{eq:eePressure1}) can be rewritten as
\beq\label{eq:eePressure2}
P_{\rm z, e e^+}&\simeq& 3.5\times 10^{23}\,b \\
&&\times \sum\limits_{n=0}^{\infty}g_n 
\int\limits_{0}^{\infty}\d p_{\rm z}\,
\frac{p^2_{\rm z}\left[ f_{-}(p_{\rm z})+ f_{+}(p_{\rm z})\right]}{(1+p_{\rm z}^2+2bn)^{1/2}} 
\,\, {\rm dyn\,\,cm^{-2}}. \nonumber
\eeq
The results of calculations with equation (\ref{eq:eePressure2}) are given in Fig.\,\ref{pic:sc_Pressure01}.
At low temperatures, when the effects of pair creation are negligible and $P\propto T$, then the pressure increases rapidly with the increase of the number density of electron-positron pairs. 
In the limiting case of high temperatures, $P\propto T^4$ and the gas pressure becomes very close to the radiation pressure, which is given in the equilibrium by
\beq\label{eq:RadPressure}
P_{\rm rad}=\frac{1}{3}a T^4\simeq 4.6\times 10^{13}\,T_{\rm keV}^4\quad {\rm dyn\,\,cm^{-2}},
\eeq 
where $a$ is the radiative constant. 

\subsection{The Eddington flux affected by electron-positron pairs}

\begin{figure}
\centering 
\includegraphics[width=8.2cm]{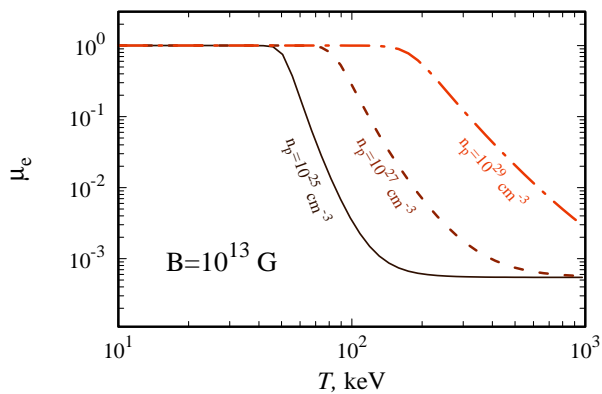} 	
\caption{The mean molecular weight as a function of temperature
for different number densities of protons: $10^{25}$ (black solid line), $10^{27}$ (brown dashed lile) and $10^{29}\,{\rm cm^{-3}}$ (red dashed-dotted line) at magnetic field strength fixed at $10^{13}\,{\rm G}$.}
\label{pic:sc_EddF}
\end{figure}

The Eddington flux is the photon energy flux which is high enough to compensate the gravitational attraction of the central object. 
The Eddington flux plays a key role in the theory of super-Eddington accretion onto compact objects and particularly in the theory of the accretion column in bright XRPs \citep{1976MNRAS.175..395B,1981A&A....93..255W,2015MNRAS.454.2539M}.
The Eddington flux is given by
\beq
F_{\rm Edd}=\frac{\rho c}{(n_{-}+n_{+})\sigma_{\rm eff}}\frac{GM}{r^2}
=\mu_{\rm e}\frac{m_{\rm p}c}{\sigma_{\rm eff}}\frac{GM}{r^2}, 
\eeq
where $\sigma_{\rm eff}$ is the effective cross section, which is affected by a strong magnetic field and local energy spectrum of photons, and the mean molecular weight is
\beq
\mu_{\rm e}\simeq\frac{1}{1836}\frac{1837 n_{\rm p}+2n_{\rm +}}{n_{\rm p}+2n_{\rm +}}.
\eeq
The temperature increase and the corresponding increase of electron/positron number density result in a drop of the mean molecular weight (see Fig. \ref{pic:sc_EddF}). 
This affects the local Eddington flux and can principally make it smaller by more than 3 orders of magnitude.
The accurate calculations of the Eddington flux require accounting for special features of Compton scattering in a strong magnetic field (see e.g. \citealt{1979PhRvD..19.2868H,1986ApJ...309..362D,2016PhRvD..93j5003M}), which is behind the scape of this letter.

\section{Summary and discussion}

We have investigated properties of electron-positron gas in accretion columns of bright X-ray pulsars accounting for specific physical conditions of a strong magnetic field and high temperatures. 
Because of a large optical thickness of accretion column due to Compton scattering, the gas of electron-positron pairs can be considered to be close to the thermodynamic equilibrium in the central parts of accretion column at high mass accretion rates typical for recently discovered ULX pulsars. However, in the outer parts accretion column the gas of pairs can be far from the equilibrium and its conditions has to be obtained from kinetic equations accounting for a number of processes (see e.g. \citealt{1985MNRAS.212..523G}).

We have demonstrated that the process of pair creation might have key importance in extreme accretion onto magnetized NSs influencing the dynamical and temperature structure of accretion columns. 
In the case of high velocity of the accretion flow and relatively low mass density, the number density of pairs becomes comparable to the number density of protons at temperatures of about a few tens of keV already (see Fig.\,\ref{pic:sc_Ne01}).
Because the energy budget of the accretion process is limited, the process of pair creation puts an upper limit on the internal temperature of the accretion column (see Fig.\,\ref{pic:sc_Tmax}) and, therefore, on the energy losses due to neutrino production \citep{2018MNRAS.476.2867M}. 
The temperatures of a few hundred keV are achievable only in the case of the sufficiently high mass density of material in the accretion channel and only in the very vicinity of a NS surface. 
The electron-positron pairs affect the gas pressure, which becomes comparable to the radiation pressure at temperatures $\gtrsim 100\,{\rm keV}$ (see Fig.\,\ref{pic:sc_Pressure01}).
The process of pair creation reduces the mean molecular weight of the accreting material (see Fig.\,\ref{pic:sc_EddF})  and, thus, drops the Eddington flux, which supports accretion columns. 
According to our knowledge, these effects were largely neglected in the models of accretion columns developed up to date and have to be taken into account in further numerical models.



\section*{Acknowledgements}

AAM and ISO acknowledge support by the Russian Science Foundation Grant No. 18-72-10070.
This work was also supported by the Netherlands Organization for Scientific Research Veni Fellowship (AAM).
We are grateful to Alexander Kaminker, Vitaly Grigoriev and an anonymous referee for discussion and useful comments.


{

}

\bsp	
\label{lastpage}
\end{document}